\documentclass[11pt]{article}
\usepackage[scale=.675,centering]{geometry}
\usepackage{amssymb,amsmath}
\usepackage{latexsym}
\usepackage{natbib}
\usepackage{fleqn,latexsym}
\usepackage{StyleFiles/smallsec,StyleFiles/nustyle2}

\renewcommand{\baselinestretch}{1.23}
\usepackage{times}

\title{Elementary Proof of a Theorem of Jean Ville%
\thanks{Thanks to Glenn Shafer for critical comment. Contact:
lieb@princeton.edu,
osherson@princeton.edu, weinstein@cis.upenn.edu.
Postal mail: Lieb, Physics, Princeton University, Princeton
NJ 08540.
Research supported by NSF grant PHY-0139984-A03
to Lieb.}}

\author{%
Elliott H.\ Lieb \\ Princeton University \and
Daniel Osherson \\ Princeton University \and
Scott Weinstein \\ University of Pennsylvania}

\newcommand{\ZETA}{\mbox{$\zeta$}}
\newcommand{\N}{\mbox{$\mathbb{N}$}}
\newcommand{\ST}{$\cal S$}
\newcommand{\QN}[1]{q\|#1}

\newcommand{\AS}{\mbox{$A^*$}}
\newcommand{\ASF}{\mbox{$C$}}

\newcommand{\BINARY}{\mbox{\textsf{B}}}

\newcommand{\PLFNC}{\mbox{${\cal E}$}}

\newcommand{\CARE}{\mbox{\textsl{care}}}
\newcommand{\CARD}{\mbox{\textrm{card}}}
\newcommand{\DISDAIN}{\mbox{\textsl{don't care}}}

\newcommand{\MID}{\mbox{$\ :\ $}}

\newcommand{\II}{\mbox{$I$}}
\newcommand{\IIA}[1]{\mbox{$\II(#1)$}}
\newcommand{\OB}{\textbf}

\newtheorem{thm}[equation]{Theorem: }

\newtheorem{defn}[equation]{Definition: }

\newtheorem{dispar}[equation]{}

\newtheorem{ex}[equation]{Example: }

\begin{document}
\maketitle

\section{Ville's Theorem}

Consider the infinite sequences of $0$'s and $1$'s, often
called \OB{reals}. Some of
them are sufficiently ``disorderly'' and ``balanced'' between $1$
and $0$ to represent the result of tossing a fair coin repeatedly,
each trial independent of the others. The remaining reals look
``fixed'' in some way, not generated \textit{randomly}. Motivating a
precise account of this distinction could elucidate fundamental
ideas in probability and statistics. \cite{Li} offer a masterful
overview of work along these lines, the earliest of which appears to be due to
Richard von Mises (1919).\nocite{Mises} To state his proposal, we
introduce some notation.

Define $\N = \{1, 2, 3, \cdots\}$, and let $n\in\N$ and real $q$ be given.
We denote the $n$th bit in $q$ by $q(n)$. The
initial finite sequence of length $n-1$ in $q$ is denoted by $q[n]$.
That is, $q[n]$ is the initial segment of $q$ that precedes $q(n)$.
For example,
if $q = 10101010 \cdots$ then $q(1) = 1$, $q[1]$ is the empty sequence which we
denote by $e$; $q[3] = 10$ and $q(3) = 1$.
The set of \textit{finite} sequences over $\{0, 1\}$ is denoted
\BINARY. A \OB{selection function} is any map of \BINARY\ into the
set $\{\CARE, \DISDAIN\}$.
Given
a selection function
$f$, the \OB{subsequence of $q$ that $f$ cares about} is
determined by including $q(n)$ in the subsequence iff $f(q[n]) =
\CARE$.
We use $\ST(q[n])$ to denote the
sum of the first $n-1$ bits in $q$. Suppose that the subsequence of $q$
that selection function $f$ cares about is infinite. Then we use
$\ST_f(\QN{n})$ to denote the sum of the first $n$ bits in this subsequence.
In other words,
\[{\cal S}_f(\QN{n}) = \sum_{k=1}^n   q(j_k)\]
where $j_1, j_2, \ldots$ are the integers $i$ such that $f(q[i])
= \CARE$. Of course, the subsequence of $q$ that $f$ cares about may
be finite or infinite.

Von Mises' idea was that some \textit{countable} collection \PLFNC\ of selection
functions would justify the following definition.
\begin{defn}\label{vonMises1}
A real $q$ is \textit{random} just in case:
\begin{enumerate}
\item\label{vonMises1a} $\lim_{n \to \infty} \ST(q[n])/n = 1/2$;
\item\label{vonMises1b} for every $f\in\PLFNC$, if
the subsequence of $q$ that $f$ cares about is infinite then
$\lim_{n \to \infty} \ST_f(\QN{n})/n = 1/2$.
\end{enumerate}
\end{defn}
Intuitively, a random real defeats any strategy of betting a fixed
stake on coordinates that are chosen by study of preceding bits. But
which countable collection \PLFNC\ of selection functions renders
\ref{vonMises1} correct, and how could this fact be demonstrated?
\cite{Lam1} and \citet[\S 1.9]{Li} review the discussion that lasted
beyond mid-century. The debate
included a striking objection to von Mises' definition that was formulated
by the French mathematician Jean Ville.
He showed that
\textit{any} choice of \PLFNC\ leads Definition \ref{vonMises1} to declare some
intuitively non-random reals to be random. Specifically:

\begin{thm}\citep{Ville}\label{satthm}
Let \PLFNC\ be any countable collection of selection functions. Then
there is a real $q$ such that:
\begin{enumerate}
\item\label{satthmc}
$\lim_{n\to\infty}\ST(q[n])/n = 1/2$.
\item\label{satthma} for every $f\in\PLFNC$, if
the subsequence of $q$ that $f$ cares about is infinite then
$\lim_{n \to \infty} \ST_f(\QN{n})/n = 1/2$.
\item\label{satthmb} for all $n \in N$, $\ST(q[n])/n \le 1/2$.
\end{enumerate}
\end{thm}
Clause (\ref{satthmb}) does the damage to von Mises' theory inasmuch
as no real $q$ that satisfies
\begin{quote}
for all $n\in N$, the number of $1$'s in
$q[n]$ does not exceed the number of $0$'s
\end{quote}
appears to be the result of independent, fair coin tosses. Indeed,
such a real falls outside of sets of measure $1$ widely
believed to hold the genuinely random sequences, e.g., those satisfying
the law of the iterated logarithm \citep[p. 157]{Feller},
and even the principle that fluctuations should be symmetrical and of order $\sqrt{n}$.

Ville's proof of \ref{satthm} is arduous, but a more compact
argument is given in \citet[pp.\ 174-6]{Shen} [relying in turn
on \citet{Loveland}]. We here exploit the
combinatorial trick introduced in the latter paper but for a
somewhat different construction (perhaps easier to follow). Both
proofs strengthen Ville's original result by showing that each
selection function in \PLFNC\ that cares about an infinite
subsequence of the constructed $q$ behaves too regularly; see
Section \ref{improv}.\footnote{See \cite{Lam1, Lam2} for proofs of
versions of the theorem, relying on probabilistic constructions.
\cite{Lam2} also discusses whether Ville's theorem is as devastating
to von Mises' program as generally believed.}

We conclude this section with some more notation.
Infinite sequences (over any set of
objects) are assumed to be ordered like \N.
Given such an infinite sequence $\gamma$, we interpret
$\gamma(n)$ and $\gamma[n]$ respectively as the contents of the
$n$th position in $\gamma$ and the initial sequence of length
$n-1$ in $\gamma$ (just as for reals).
A \OB{tail} of an
infinite sequence $\gamma$ is any subsequence of $\gamma$ that
excludes just a finite initial segment.
Given two finite sequences $\tau, \sigma$ over any set of objects,
the concatenation of $\tau$ to the end of $\sigma$ is denoted
$\sigma\tau$. We'll make use of the following example.

\begin{ex}\label{fzero}
One selection function, $h$, satisfies:
\[h(\sigma) = \CARE \mbox{ for all }\sigma\in\BINARY.\]
\end{ex}
Thus, for all reals $q$, the subsequence of $q$ that $h$ cares about is all of $q$.

\section{Intuitive motivation for the proof}

We attempt to convey the underlying idea of our proof of Theorem
\ref{satthm}.
Subsequent developments are self-contained, so the present
section may be skipped.
Let us first consider a weaker version of Ville's theorem, in which
\PLFNC\ is \textit{finite}.
\begin{dispar}\label{satthmFinite}\textsc{Finite version of Ville's Theorem}:
Let \PLFNC\ be any finite collection of selection functions. Then
there is a real $q$ such that:
\begin{enumerate}
\item\label{satthmFinitec}
$\lim_{n\to\infty}\ST(q[n])/n = 1/2$.
\item\label{satthmFinitea} for every $f\in\PLFNC$, if
the subsequence of $q$ that $f$ cares about is infinite then
$\lim_{n \to \infty} \ST_f(\QN{n})/n = 1/2$.
\item\label{satthmFiniteb} for all $n\in\N$, $\ST(q[n])/n \le 1/2$.
\end{enumerate}\end{dispar}
To prove \ref{satthmFinite}, we shall assume that $h$ of Example \ref{fzero} is a member of \PLFNC.
Then it suffices to construct a real $q$ that satisfies clauses
(\ref{satthmFinitea}) and (\ref{satthmFiniteb}).
We construct the desired $q$ in stages, $q(1), q(2), \ldots$.
At each stage $n$, we also define the subset $\ASF(n)$ of \PLFNC\ that cares
about $q[n]$.
\begin{quote}
\underline{Stage $n$}: Suppose that $\ASF(m)$ for all $m < n$ and $q[n]$ have been defined.
Set $\ASF(n) = \{f\in\PLFNC\MID f(q[n]) = \CARE\}$. Set
$q(n) = \CARD\{j < n\MID \ASF(j) = \ASF(n)\}\mod 2$.
\end{quote}
In words, we set the bit $q(n)$ to zero if the subset of \PLFNC\
that cares about the initial segment of length $n-1$ [namely, $\{f\in\PLFNC\MID f(q[n]) =
\CARE\}$] appears an even number of times earlier in the
construction; otherwise, we set $q(n)$ to one. It is obvious that
$q$ satisfies \ref{satthmFinite}\ref{satthmFiniteb} since every $1$
appearing in $q$ is preceded by an occurrence of $0$ that can be
uniquely chosen to match it.

Let $f\in\PLFNC$ be given with $\{n\MID f(q[n]) = \CARE\}$ infinite.
(If there are no such $f$ in \PLFNC, we are done.) Let $n_1, n_2, \ldots$ be an
increasing enumeration of $\{n\MID f(q[n]) = \CARE\}$. Then $B =
\ASF(n_1), \ASF(n_2), \ldots$ contains exactly the members of the sequence
\ASF\ that include $f$, in particular, no set appearing in $B$ also
appears outside of $B$. Hence, for all $m\in\N$, the value of
$q(n_m)$ depends on just $B$. Subsets of \PLFNC\ that occur only
finitely often in $B$ ultimately stop occurring altogether since there
are only finitely many of them. Therefore, the number of $1$'s and $0$'s in
$q[n_m]$ is ultimately governed by the subsets
of \PLFNC\ that occur infinitely often in $B$. The latter collection
is nonempty because $B$ is infinite and there are only finitely many
distinct subsets of \PLFNC\ that contain $f$ (so at least one of
them must occur infinitely often in $B$).
Observe also that for $k = \CARD(\PLFNC)$, no more than $2^k$ zeros can
occur consecutively in $q$ since a block of zeros requires that different
subsets of \PLFNC\ care about each coordinate in the block.
The construction
of $q$ now makes it evident that
\[\lim_{m\to\infty}\frac{\CARD\{j\MID q(n_j) = 1 \mbox{ and } j \le m\}}{m} = \frac12,\]
demonstrating \ref{satthmFinite}\ref{satthmFinitea}, and finishing the proof
of the finite version of Ville's theorem.
Indeed, our construction proves a little more inasmuch as it guarantees that
for every selection function $f$ with $\{n\MID f(q[n]) = \CARE\}$ infinite,
\[0 \le \frac{n}{2} - {\cal S}_f(\QN{n}) \le 2^{\textrm{card}(\cal E)} \mbox{\quad for all }n.\]

How can we extend this reasoning to Theorem \ref{satthm}? We can't
consider subsets of an infinite collection of selection functions
since each might occur just once in the sequence $\ASF$. This would
make $q$ into a sequence of zeros. The next idea might be to
enumerate \PLFNC\ as $f_1, f_2, \ldots$, then carry out the
foregoing construction with $\{f_i \MID i \le n\}$ for increasing
values of $n$. In other words, we would build a real $q$ as in the
finite case for $\{f_1\}$ but stop at $q[k_1]$ for $k_1$ large
enough to ensure that $\ST_{f_1}(\QN{n})/n$ is at least $1/4$, where
$n$ is the number of bits in $q[k_1]$ that $f_1$ cares about. Then
we would continue to build $q$ starting at $q[k_1]$ but this time on
the basis of $\{f_1, f_2\}$. We would stop at $q[k_2]$ for $k_2 >
k_1$ large enough to ensure that both $\ST_{f_1}(\QN{m})/m$ and
$\ST_{f_2}(\QN{n})/n$ are at least $3/8$, where $m$ and $n$ are the
numbers of bits in $q[k_2]$ that $f_1$ and $f_2$ care about,
respectively. And so forth.

This seductive plan is foiled, however, by the prospect that $f_2$,
for example, will cease to care about $q$ prematurely during the
second stage, making it impossible to ensure that
$\ST_{f_2}(\QN{n})/n \ge 3/8$. Yet if we continue the construction
despite this setback, there is no guarantee that $f_2$ will care
only finitely often in $q$ overall rendering its behavior
irrelevant. Indeed, $f_2$ might care exactly once in stage 3,
perhaps at the same initial segment as $f_3$, then care exactly once
in stage 4, perhaps at the same initial segment as $f_4$, and so
forth. In the end, $f_2$ may care infinitely often but almost always
in the context of a unique set of other selection functions. In this
case, $\ASF(k)$ will be a new subset of \PLFNC\ for cofinitely
many $k$ among $\{j\MID f_2(q[j]) = \CARE\}$. In turn, $q(k)$
will be set to zero for a cofinite subset of the coordinates where
$f_2$ cares.\footnote{Another approach is to attempt to map each
selection function $f$ into another $f^\dag$ such that for all reals $q$,
$\{i\MID f^\dag(q[i]) = \CARE\}$ is infinite, and
$\{i\MID f^\dag(q[i]) = \CARE\} = \{i\MID f(q[i]) = \CARE\}$ if the latter
set is infinite. It can be shown, however, that there is no such mapping.
\textit{Hint}: Consider the selection function that cares about $\sigma\in
\BINARY$ iff $1$ appears somewhere in $\sigma$ (i.e., $\sigma$ is not
a block of $0$'s).}

Our proof of Ville's Theorem extends the construction for
the finite case but uses a combinatoric trick to avoid the difficulty
just described. At stage $n$ of the
construction of $q$ we build a finite subset $\ASF(n)$ of \PLFNC\
that is used to determine $q(n)$ as in the finite case (by
determining the parity of the set of its previous co-occurrences in
the construction). The rule for constructing the sequence $\ASF$,
however, does not allow $f_k$ to appear with $f_{k+m+1}$ until it
has appeared sufficiently often by itself or with some of $f_1
\ldots f_{k+m}$. By defining ``sufficiently often'' in the right
way, this maneuver builds up enough parity reversals to ensure that
$\lim_{n \to \infty} \ST_{f_k}(\QN{n})/n = 1/2$ if the subsequence of
$q$ that $f_k$ cares about is infinite.

To make all this clear, it will be notationally simpler to work with just the
indexes of our selection functions. We start by presenting the combinatorial
core of the argument before turning to its application to Ville's Theorem.

\section{A combinatorial construction}\label{heart}

Let $\cal A$ be the class of infinite sequences of subsets of \N\ that contain
$1$; that is, for $A\in \cal A$ and $i\in N$,
$A(i)\subseteq \N$ and $1\in A(i)$. We define a map $^*$ from $\cal
A$ into itself. We denote the result of applying the map to
$A\in\cal A$ by \AS. For $A\in\cal A$, each coordinate of $\AS$ will
be a nonempty, finite subset of the corresponding coordinate of $A$.
To describe $^*$ let $A\in \cal A$ be given.
$A^*(n)$ will be the subset of $A(n)$
consisting of the  numbers in $A(n)$ that are less than or equal to a certain
number \IIA{n} which, in turn, will be determined by $A[n]$.

\textit{Stage $n$ of the construction of \AS}: We suppose that for all $m < n$,
$A^*(m)$ and \IIA{m} have been constructed with
\[A^*(m) = \{j\in A(m)\MID 1 \le j\le \IIA{m}\}.\]
Then we define:\begin{dispar}\label{Idisplay}
$\begin{array}{lll}
\IIA{n} & = & \min i(\exists j\in A(n)\MID \CARD\{ m < n \MID j\in A^*(m) \mbox{ and }
\IIA{m} = i\}\le 3^i)\\
A^*(n) & = & \{j \in A(n)\MID 1 \le j \le \IIA{n}\}
\end{array}$\end{dispar}
Note that $\IIA{1} = 1$ and $A^*(1) = \{1\}$. Evidentally:

\begin{dispar}\label{continuity}
The construction of $\AS(n)$ depends on just
$\{A(i)\MID i\le n\}$.
\end{dispar}

\noindent
It is also easy to see that:
\begin{dispar}\label{finFact}
For all $i \in\N$, $\IIA{n} = i$
for only finitely many $n$ (indeed, for at most $i \cdot 3^i$ many $n$).
\end{dispar}

Now fix $\ell\in\N$ and suppose that it occurs infinitely often in
$A$ (for example, $\ell$ might be $1$). Let $\{n\MID \ell\in A(n)\}$
be enumerated in increasing order as $n_1, n_2, \cdots$. Then by
\ref{finFact}:
\begin{dispar}\label{finFact2}
For cofinitely many $m\in \N$, $\ell\in \AS(n_m)$.
\end{dispar}

Now we consider the sequence of integers $\ZETA = I(n_1), I(n_2), \cdots$.
It follows at once from \ref{Idisplay} that:
%
%
\begin{dispar}\label{alphaFact1}
for all $k \ge \ell$, there are at least $3^k$ many occurrences of $k$ in
\ZETA\
prior to the first occurrence of $k+1$ in \ZETA.
\end{dispar}
\noindent

For $k\ge \ell$, define:
\[
\alpha(k) = \AS(n_m), \AS(n_{m+1}), \cdots, \AS(n_{m+r})\]
where $n_m$ is the first occurrence of $k$ in \ZETA, and
$n_{m+r+1}$ is the first occurrence of $k+1$ in \ZETA.
{}From \ref{finFact2} and \ref{alphaFact1}, we have:
\begin{dispar}\label{PartFact}
There is $k\ge \ell$ and tail $t$ of $\AS(n_1), \AS(n_2), \cdots$ such that:
\begin{enumerate}
\item\label{PartFactb} $t$ has the form
$\alpha(k)\,\alpha(k+1)\,\alpha(k+2)\, \cdots$
\item\label{PartFacta} $\ell$ is a member of every coordinate of $t$.
\end{enumerate}
\end{dispar}
\noindent
Specifically, $k$ can be chosen to be the first occurrence of a number in
\ZETA\ such that all later numbers occurring in \ZETA\ are greater than $\ell$.
Now fix some $k$ and $t$ as described in \ref{PartFact}. (We leave implicit the
dependence of $k$ and $t$ on $\ell$.)
By the definition of $n_1, n_2 \cdots$, we have:
\begin{dispar}\label{redherring1}
For cofinitely many members $m$ of $\{n\MID\ell\in A(n)\}$,
$\AS(m)$ appears in $t$.
\end{dispar}
\noindent
{}From the definition of $\alpha(i)$, for all $i \ge k$,
each of the sets appearing in
$\alpha(i)$ is a subset of $\{1 \cdots i\}$ so there
are at most $2^i$ of them. Along with \ref{alphaFact1}, this yields:
\begin{dispar}\label{PartFact2}
$t$ has the form $\alpha(k)\,\alpha(k+1)\,\alpha(k+2)\, \cdots$,
where for all $m \ge 0$, $\alpha(k+m)$ has length at least $3^{k+m}$
and contains at most $2^{k+m}$ distinct sets.
\end{dispar}

\section{From finite sets to bits}

Recall that we have fixed
$A\in {\cal A}$, and thus also fixed \AS. We describe a method for mapping \AS\ into a
real $q$. For $n\in N$, the \OB{preceding parity of $\AS(n)$ in
$\AS$} denotes:
\[
\CARD\{j < n\MID \AS(j) = \AS(n)\}\mod 2.
\]
That is, the preceding parity of $\AS(n)$ in $\AS$ is $0$ if $\AS(n)$
appears earlier in $\AS$ an even number times; it is $1$ if it
appears an odd number of times. The real $q$ is now defined as
follows. For all $n\in N$, $q(n)$ is the preceding parity of
$\AS(n)$ in $\AS$.

Let $n\in N$ be given, and consider
\[B_0 = \{ i \le n\MID q(i) = 0\}\]
\[B_1 = \{ i \le n\MID q(i) = 1\}.\]
The construction of $q$ implies that each member of $B_1$ can be paired
with a unique, smaller member of $B_0$. Therefore:

\begin{dispar}\label{villeProp}
For all $n\in\N$, $\ST(q[n])/n\le 1/2$.
\end{dispar}

Recall that we also fixed
$\ell\in \N$ that occurs in infinitely many coordinates of $A$.
As before, let $\{n\MID \ell\in A(n)\}$ be enumerated in increasing
order as $n_1, n_2, \cdots$. Let $\hat q$ denote $q(n_1), q(n_2)
\cdots$ We wish to demonstrate that:
\begin{dispar}\label{finalLem1}
$\lim_{n\to\infty}\ST(\hat q[n])/n = 1/2$.
\end{dispar}
For this purpose it suffices to exhibit a tail $s$ of $\hat q$ that:
\begin{dispar}\label{finalLemHelp1}
$\lim_{n\to\infty}\ST(s[n])/n = 1/2$.
\end{dispar}
To specify $s$, let $t$ be the tail of
$\AS(n_1), \AS(n_2), \cdots$ described in \ref{PartFact2}. We define $s$ to be
such that $s(1) = \hat q(n_m)$ iff $t(1) = \AS(n_m)$. [That is, $s$ excludes
an initial segment of $\hat q$ equal in length to the initial segment of
$\AS(n_1), \AS(n_2), \cdots$ excluded by $t$.]
We now show that
this $s$ conforms to \ref{finalLemHelp1}.

Recall from \ref{PartFact} that $t$
has the form
$\alpha(k)\,\alpha(k+1)\,\alpha(k+2)\, \cdots$, and is such
that for all $i\in N$, $\ell\in t(i)$.
Let $j \ge 0$ be given, thought of as a coordinate of $t$ and also of $s$. Without loss of
generality, we assume that $j$ is big enough so that there is $m(j)$ such
that $t(j)$ falls within $\alpha(k+m(j)+1)$. We define
\[
N_0(j) = \mbox{the number of $0$'s in $s[j]$, and}
\]
\[
N_1(j) = \mbox{the number of $1$'s in $s[j]$}.
\]
There follow some properties of $N_0(j)$ and $N_1(j)$ which are consequences of
\ref{PartFact2} and the fact that $t$ is composed of all and only the sets
of \AS\ that contain $\ell$, except for a finite ``head.''
[The preceding parity of $t(j)$ in \AS\ therefore depends
on just the preceding members of $t$.]

First, since the block $\alpha(k+m)$ has at least
$3^{k+m}$ coordinates, we have:
\begin{dispar}\label{Nprops}
$N_0(j) + N_1(j) \ge 3^{k+m(j)}$.
\end{dispar}
{}From \ref{PartFact2}, there are at most $2^{k+i}$ distinct sets
in $\alpha(k+i)$, and this number bounds the number of unmatched $0$'s. So:
\begin{dispar}\label{complexThing}
\[N_0(j) \le N_1(j) + \sum_{i=0}^{m(j)+1} 2^{k+i} \le N_1(j) + 2^{k+m(j)+2}.\]
\end{dispar}
{}From \ref{complexThing} we infer:
\begin{dispar}\label{complexThing2}
\[N_1(j) \ge \frac12\left(N_0(j) + N_1(j) - 2^{k+m(j)+2}\right).\]
\end{dispar}
Let $p$ be the length of the ``head'' missing from $s$. Then:
\begin{dispar}\label{complexThing3}
\[N_1(j) \le N_0(j) + p.\]
\end{dispar}
This inequality allows for the presence of unmatched $0$'s in the head,
which would induce unmatched $1$'s afterwards.
Similarly to the transition from \ref{complexThing} to \ref{complexThing2},
we see that \ref{complexThing3} implies:
\begin{dispar}\label{complexThing4}
\[N_1(j) \le \frac12(N_0(j) + N_1(j) + p).\]
\end{dispar}
We now evaluate $R(j) = N_1(j)/(N_0(j) + N_1(j)$. Because we've neglected only
finitely many terms [that is, $R(j)$ for $j$ with $t(j)$ a coordinate of
$\alpha(k)$], it is clear that if $\lim_{j\to \infty} R(j) = 1/2$ then
\ref{finalLemHelp1} is true. For an upper bound, we use \ref{complexThing4} and compute:
\[
R(j) \le \frac{N_0(j) + N_1(j) + p}{2(N_0(j) + N_1(j))}
\]
which goes to $1/2$ as $j$ goes to infinity.
For the lower bound, we use \ref{complexThing2} and calculate:
\[
R(j) \ge \frac{N_0(j) + N_1(j) - 2^{k+m(j)+2}}{2(N_0(j) + N_1(j))} = \frac12 - \frac{2^{k+m(j)+1}}{N_0(j) + N_1(j)},
\]
and this also converges to $1/2$ in view of \ref{Nprops}.

\section{Application to Ville's theorem}\label{appV}

To return to Ville's Theorem \ref{satthm}, without loss of
generality we may assume that \PLFNC\ can be enumerated without
repetition as $f_1, f_2 \cdots$ where $f_1$ is the ``always care''
function of Example \ref{fzero}. For, it's clear that if
\ref{satthm} holds for $\PLFNC'\supseteq\PLFNC$ then it holds for
\PLFNC. So, in the preceding construction, we may conceive of the
members of $A(i)$ --- the coordinates of the infinite sequence of
subsets of $N$ --- as indexes for selection functions in \PLFNC. Our
goal is to construct a real $q = q(1), q(2), \ldots$ with the
properties stated in Theorem \ref{satthm}. Because the ``always care''
function appears in \PLFNC, it suffices to demonstrate
\ref{satthm}\ref{satthma},\ref{satthmb}.

The construction is built on the results of the previous
sections. There, we were given an infinite sequence $A(1), A(2), \ldots$
of subsets of ${\Bbb{N}}$ and these were reduced, by our construction,
to an infinite sequence $\AS(1), \AS(2), \ldots$ of finite subsets of
${\N}$. [In fact, $\AS(n) \subseteq A(n) $ for all $n$.]  Finally,
we showed how to map \AS\ into a real $q(1), q(2), \ldots$.

We note that  the value of $q(n)$ depends only on $A[n+1] = \{A(1),
A(2), \ldots, A(n)\}$.  Therefore, all we have to do for Ville's theorem
is to start with $A(1) = \{m\in\N\MID f_m(e) = \CARE\}$, and produce
$q(1)$ on the basis of $\AS(1)$. [It's easy to see that $q(1) = 0$.]
Next we define $A(2) = \{m\in\N\MID f_m(q(1)) = \CARE\}$, and
produce $q(2)$ from $\AS(1), \AS(2)$. Similarly,
$A(3)$ is the subset of \N\
consisting of the subscripts of all selection functions
that care about the finite sequence $q(1), \ q(2)$, and
so on, ad infinitum.

The real $q$ that witnesses Ville's theorem
has now been constructed. The bounds \ref{complexThing2}, \ref{complexThing3}
describe the number of $1$'s and $0$'s that appear in the subsequence of
$q$ about which $f_\ell$ ``cares.''
This concludes the proof of Ville's theorem in its original
formulation. In other words, we have constructed a binary sequence with
the property that the entire sequence has a running sum $S_1(n) $ that
never exceeds $n/2$ and yet each selection function $f_\ell$ that cares infinitely
often has a ratio
$S_\ell(n) /n$ that converges to $1/2 $ as $n\to \infty$. But much more
can be learned from \ref{complexThing2}, \ref{complexThing3} that were not
previously noted, as far as we are aware.

\section{Improvements to Ville's Theorem}\label{improv}

Let $q$ be the real constructed by the method described above.
Choose a selection function $f_\ell$ that ``cares'' about $q$ infinitely often
(e.g., $f_1$).
We define the \OB{fluctuation} (or fluctuation about the mean) for
selection function $f_{\ell}$ to be
\[
\delta_{\ell }(n) =  S_{f_\ell}(\QN{n}) - n/2.\]
{}From \ref{complexThing3} we learn that $\delta_{\ell}$
is bounded above by an $\ell$-dependent
constant. This property mimics the behavior of the fluctuation for
the entire $q$ sequence (i.e., for $f_1$), whose fluctuation is never
positive.

For a bound in the other direction, we can use \ref{Nprops} and \ref{complexThing2}
to conclude that there is a number
$C_{\ell} \geq 0$ such that for all $n$
\begin{dispar}\label{log3}
$\delta_{\ell} (n) \geq -C_{\ell} \ n^{\ln 2/
\ln 3}$.
\end{dispar}
\noindent A quick look at our proof, however, shows that the
appearance of $\ln 3$ in \ref{log3} comes from our use of $3^i$ in
the definition \ref{Idisplay} of $I(n)$. We could have used $r^i$
instead, as long as $r > 2$, notably, $r = 2^{1/\varepsilon}$ with
$\varepsilon < 1$. By replacing the number $3$ by $r$ in the
preceding sections, and making no other changes, we conclude that for
every $\varepsilon > 0$, there is a constant $C_{\ell} (\varepsilon)
\geq 0$ such that:
\begin{dispar} \label{epsilon}
for every $n$,
$\delta_{\ell} (n) \geq -C_{\ell} (\varepsilon) \ n^{\varepsilon}$.
\end{dispar}
The existence of an $n$-independent upper bound is not
affected by this change of $3^i $ to $r^i$.

The bound \ref{epsilon} is indeed remarkable. For random coin
tosses the law of the iterated logarithm states that the fluctuations exceed $(1  -
\varepsilon')\sqrt{n\ln\ln n}/\sqrt{2}$ (for any $\varepsilon' > 0$)
infinitely often almost surely \citep{Feller}.
Our fluctuations are absolute, not
probabilistic, and suggest that a more clever strategy would reduce
the fluctuations even further. Indeed, it is easy to see that for
any slow-growing function $g,$ for example $\ln n,$ there is a
suitably fast-growing function $h,$ so that our construction with
$h(i)$ in place of $3^i$ will enforce a bound analogous to
\ref{epsilon} with $g(n)$ in place of $n^{\varepsilon}$ and a
constant  $C_{\ell} (g)$ in place of $C_{\ell} (\varepsilon)$.

\renewcommand{\baselinestretch}{1.0}
\bibliography{cohbib2}
\bibliographystyle{StyleFiles/econometrica}
\end{document}